\documentclass[english,aps,pra,reprint,showpacs,titlepage,longbibliography]{revtex4-1}   

\usepackage[T1]{fontenc}	
\usepackage[latin9]{inputenc}	
\usepackage{geometry}		
\usepackage{graphicx}
\usepackage[above,below]{placeins}	
\usepackage{times}

\usepackage{hyperref}  
\hypersetup{colorlinks=true,urlcolor=blue,citecolor=blue,linkcolor=blue}   
\usepackage{enumitem}          
\setlist{nosep}                 

\usepackage[dvipsnames]{xcolor}
\newcommand{\drdf}[1]{#1}
\newcommand{\redact}[1]{#1} 

\begin{document}

\begin{titlepage}

\title{Student perceptions of laboratory classroom {activities} and experimental physics practice}

\author{Dimitri R. Dounas-Frazer}
\affiliation{Department of Physics and Astronomy, Western Washington University, Bellingham, WA 98225, USA}
\affiliation{SMATE, Western Washington University, Bellingham, WA 98225, USA}

\author{Kimme S. Johnson}
\affiliation{Woodring College of Education, Western Washington University, Bellingham, WA 98225, USA}

\author{Soojin E. Park}
\affiliation{Department of Anthropology, Western Washington University, Bellingham, WA 98225, USA}
\affiliation{{Woodring College of Education, Western Washington University, Bellingham, WA 98225, USA}}

\author{Jacob T. Stanley}
\affiliation{BioFrontiers Institute, University of Colorado Boulder, Boulder, CO 80309, USA}

\author{H. J. Lewandowski}
\affiliation{Department of Physics, University of Colorado Boulder, Boulder, CO 80309, USA}
\affiliation{JILA, National Institute of Standards and Technology and University of Colorado Boulder, Boulder, CO 80309, USA}

\begin{abstract}
{
We report results from a study designed to identify links between undergraduate students' views about experimental physics and their engagement in multiweek projects in lab courses. Using surveys and interviews, we explored whether students perceived particular classroom activities to be features of experimental physics practice. We focused on 18 activities, including maintaining lab notebooks, fabricating parts, and asking others for help. Interviewees identified activities related to project execution as intrinsic to experimental physics practice based on high prevalence of those activities in interviewees' own projects. Fabrication-oriented activities were identified as conditional features of experimentation based on differences between projects, which interviewees attributed to variations in project resources. Interpersonal activities were also viewed as conditional features of experimentation, dependent upon one's status as novice or expert. Our findings suggest that students' views about experimental physics are shaped by firsthand experiences of their own projects and secondhand experiences of those of others.
}\clearpage
\end{abstract}

\maketitle

\end{titlepage}

\section{Introduction and background}
Undergraduate physics lab courses are characterized by a variety of learning objectives~\cite{AAPT2014}, including developing proficiency with troubleshooting~\cite{Dounas-Frazer2017a}, modeling~\cite{Dounas-Frazer2018c}, and technical writing~\cite{Hoehn2020,Stanley2018,Stanley2016b,Ramey2018,Coppens2016}. {National} interview studies with instructors have found that some instructors want students to develop sophisticated views about experimental physics, such as the belief that `nothing works the first time' when conducting experiments~\cite{Dounas-Frazer2016} or the perception of experimentation as an iterative process~\cite{Dounas-Frazer2017c}. {Skills- and views-based learning objectives may be intertwined~\cite{Wilcox2017a}, suggesting that student engagement in experimentation can simultaneously develop students' proficiency with physics skills while shaping their perceptions about physics practice}. Relatedly, the {goal} of this paper is to identify links between students' engagement in multiweek final projects and their views about experimental physics.

For several years, researchers have used the Colorado Learning Attitudes about Science Survey for Experimental Physics (E-CLASS) to gain insight into students' views about experimental physics. Developed by Zwickl et al.~\cite{Zwickl2014}, E-CLASS is a Likert-style survey that probes students' ideas about what experimentation entails and their perceptions of experimental physics as doable or enjoyable. {Analysis} of E-CLASS scores shows that students' views are different from those of practicing experimental physicists~\cite{Wilcox2017b}. In {an investigation} of students' rationale for their responses to E-CLASS items, Hu et al.~\cite{Hu2017} found that students' views can be negatively impacted by their engagement in highly guided lab activities {in which} instructions can be followed without understanding the relevant physics concepts. Consistent with the findings of Hu et al., Wilcox and Lewandowski~\cite{Wilcox2016b} demonstrated that students' post-instruction E-CLASS scores are more consistent with {expert-like} responses in courses that incorporate open-ended activities compared to those that do not. To explain their results, Wilcox and Lewandowski hypothesized that ``open-ended activities may provide greater opportunities for the students to engage authentically in the process of experimental physics'' (p.\ 020132-6).

Wilcox and Lewandowski's hypothesis has support in the physics education \drdf{literature. Irving} and Sayre investigated the experiences of students in an advanced lab course with multiweek experiments. In the advanced lab course, students worked on long and difficult experiments during which they engaged in activities that align with the authentic practice of physics. Irving and Sayre argued the course simulated the experiences of participating in a practicing community of physicists, thus supporting students to develop knowledge about physics practice~\cite{Irving2014}. Quan and Elby studied the experiences of students working on {semester-long} projects in a research course for first-year physics majors. In the research course, projects spanned theoretical and experimental topics, and students participated in a regular seminar in which they reflected on their experiences working on projects. Quan and Elby showed that, for some students, participating in authentic research can lead to coupled shifts in their \drdf{confidence} and their view of science as an endeavor in which novices can make meaningful contributions~\cite{Quan2016b}.

Additional research is needed to fully explore the landscape of mechanisms that explain how students' views about experimental physics are shaped by their {engagement in authentic experimentation}. Here, we describe a study that complements prior work~\cite{Wilcox2016b,Irving2014,Quan2016b} by examining students' perceptions of the authenticity of specific activities during the final project portion of an upper-level lab course.

In our study, we align our definition of experimental physics with Ford's~\cite{Ford2015} definitions of scientific performance and practice. Drawing on the Next Generation Science Standards~\cite{NGSS2013} and philosophical work by Rouse~\cite{Rouse2007}, Ford defines performances as the constituent activities of scientific practice, and scientific practice as a set of connected performances whose collective purpose is to explain nature better~\cite{Ford2015}. Ford's work is a useful lens through which to investigate physics projects{, as demonstrated by Quan et al.~\cite{Quan2018}.}

Following Ford's lead, we view experimental physics practice as comprised of connected performances, such as building apparatus or analyzing data, whose purpose is to explain the physical world better. Although Ford's notions of performance and practice did not inform our study design, they provide language for articulating our research questions:
\begin{enumerate}
\item[Q1.] When completing projects in a lab course, which activities do students perceive to be constituent performances of experimental physics practice, and why?
\item[Q2.] How does participation in projects inform students' ideas about what experimental physics practice entails?
\end{enumerate}

\vspace{-10pt}
{\section{Context, participants, and methods}}
\drdf{To probe students' thoughts about their final projects, we collected survey and interview data from undergraduate physics students enrolled in upper-level optics and lasers lab courses. The courses were required for some physics bachelor's degree tracks} at a private, selective, Christian, Predominantly White Institution {(PWI)} in the Midwestern United States. Typical enrollment in each course was about 20 students per course. Averaged over five years, 13\% of course completers were women, and 87\% were men; 6\% were students of color, and 94\% were white \footnote{We report gender- and race-based demographics for three reasons: to avoid over-generalizing our results, to challenge tacit assumptions that physics students are white men unless otherwise specified, and to enable meta-studies of the physics education literature.}.

The optics and lasers lab courses were similar to one another in content and format. Learning objectives included developing students' competence with optics- and lasers-related topics and skills. Each course was divided into two seven-week halves. In the first half, students completed weekly guided lab activities. In the second half, they worked in groups {of two to four students} to complete projects, such as building a plasmon laser or achieving single-photon interference. Groups were assigned based {on} students' shared interest in a topic. Projects culminated in written reports and oral presentations. \redact{We have previously analyzed data from this population in other studies~\cite{Stanley2016, Dounas-Frazer2017b}, and a detailed description of the {course contexts} can be found in Ref.~\cite{Dounas-Frazer2017b}.}


\begin{table*}[t]
\vspace{-6pt}
\caption{\drdf{Activities emerged from analysis of student responses to weekly reflection prompts. The degree to which students perceived each activity to be a feature of experimental physics was probed during a post-project Likert-style survey, and students' rationale for survey responses was probed in follow-up interviews. Columns represent groupings that emerged during analysis of survey and interview data.}\label{table:performances}}
\begin{ruledtabular}
\begin{tabular}{*{5}{l}}
Execution-oriented & Interpersonal & Fabrication-oriented & Propagation-oriented \\ \hline
Setting up equipment & Asking a supervisor for help & Fabricating parts and materials & Reading scientific papers \\
Troubleshooting problems & Asking peers for help & Building electronics & Reading technical data sheets \\
Maintaining a lab notebook & Confirming results with an expert & Writing code to interface with equipment & Writing lab reports \\
Analyzing data & Dividing labor among team members & Writing code to simulate results & Presenting results orally \\
& Making decisions as a team & & Reflecting on progress
\end{tabular}
\end{ruledtabular}
\vspace{-6pt}
\end{table*}

Data collection was led by \redact{authors DRDF, JTS, and HJL.} We collected data from one instance each of the optics and lasers courses. {Out of 36 total students, 35} agreed to participate in our study. 
Demographics of research participants closely match those of course completers.

We collected data using free-response surveys, a Likert-style survey, and post-instruction interviews. While projects were ongoing, we administered weekly free-response surveys that prompted participants to reflect on their goals, challenges, and {successes}; \redact{for more details, see Ref.~\cite{Dounas-Frazer2017b}}. \drdf{Each week, authors DRDF and JTS read through student reflections and collaboratively generated a summary of students' progress on their projects. Through this process, we identified 18 activities that were common topics of reflection (Table~\ref{table:performances}).}

After identifying the activities in Table~\ref{table:performances}, we created a three-point Likert-style survey that prompted participants to evaluate the extent to which each activity is a feature of experimental physics research: \emph{not a feature}, \emph{might be a feature}, or \emph{definitely a feature of experimental physics research}. The Likert-style survey was administered at the end of the semester, and post-project interviews were conducted shortly thereafter. Interviews with students in the optics lab were conducted in person by \redact{HJL}, and interviews with students in the lasers lab were conducted remotely by \redact{DRDF.} During interviews, the interviewer asked participants to explain their rationale for each response on the Likert-style survey.

Analysis of interview transcripts was led by \redact{authors DRDF, KSJ, and SEP in consultation with HJL}. The unit of analysis was a participant's explanation for their response to a single item on the Likert-style survey. {Our goal was to characterize which activities were perceived as features of experimental physics, and why.} We engaged in two iterations of collaborative coding during which we coded 630 transcript excerpts. Collaborative coding {consisted of} multiple coders simultaneously evaluating a transcript excerpt and reaching consensus on an interpretation. Throughout both iterations, we regularly discussed our methods and interpretations with other physics education researchers~\redact{\footnote{Participants in weekly dialogues included B.\ Pollard, C.\ L.\ Ramey II, and L.\ R\'ios; B.\ R.\ Wilcox and D.\ Hu; A.\ Elby and G.\ M.\ Quan; A.\ Gupta and C.\ Turpen; A.\ J.\ Little, C.\ Alvarado, and S.\ A.\ Hyater-Adams; and J.\ J.\ Chini, E.\ Scanlon, and B.\ Zamarripa Roman}} in a deliberate effort to incorporate dialogue into our process for generating claims~\cite{HillCollins1989}.

In the first iteration of coding, \redact{KSJ and SEP} collaboratively identified emergent themes and developed an initial codebook consisting of code definitions, inclusion criteria, and exemplars~\cite{Saldana2015}. \redact{DRDF} played a supervisory role. Preliminary results were presented at a conference~\cite{Johnson2019,Park2019}, after which we reflected on questions and suggestions that arose during the presentations. In the second iteration of coding, \redact{DRDF, KSJ, and SEP} revised the codebook by redefining some emergent codes. {Selected} codenames, definitions, and exemplars from the revised codebook are listed here:
\begin{itemize}
\item \emph{Necessary:} interviewee referred to an activity as an inevitable, necessary, or required part of research. ``You're going to take data if you're doing research, so then you'd have to analyze it for it to be useful.''
\item \emph{Not necessary:} interviewee referred to an activity as important, helpful, or common, but not a necessary part of research. ``You don't always have to give [oral presentations], but it's commonly something to do.''
\item \emph{Status-dependent:} interviewee refers to their own or others' status as a novice or expert experimentalist. ``You might be the expert \ldots People could check your work, but I don't know necessarily that they'll be any better at understanding it than you will be.''
\item \emph{Resource-dependent:} interviewee referred to availability of resources. ``We had to build a lot of them [electronics] here to try to save money, but if you have the money, it's nice not to have to build everything.''
\item \emph{Secondhand experience:} interviewee referred to others' experiences in the course or related contexts. ``I didn't have to do that [write code to simulate results], but I know other groups that had to do that.''
\item \emph{Firsthand experience:} interviewee referred to their own experiences in the course or related contexts. ``We had to fabricate our diode laser gain material.''
\end{itemize}
The revised codebook differed from the initial one in several ways. For example, although the initial codebook included codes related to students' context-dependent and experience-oriented explanations, it did not distinguish between status- and resource-dependence or between secondhand and firsthand experiences. \redact{Finally, DRDF, KSJ, and SEP} collaboratively recoded the data by applying the revised codebook to all 630 transcript excerpts, and they discussed findings and interpretations with the other coauthors.

\vspace{-10pt}
\section{Results}
\drdf{Data analysis revealed the following patterns (Table~\ref{table:performances}):}
\begin{itemize}
\item \drdf{Four activities were almost unanimously perceived as \emph{definitely a feature of experimental physics research}; because these activities relate to the execution of a research project, we labeled them \textbf{execution-oriented}.}
\item \drdf{Five activities accounted for almost all \emph{status-dependent} codes; because these activities involve interactions among people, we labeled them \textbf{interpersonal}.}
\item \drdf{Four activities accounted for almost all \emph{resource-dependent} and \emph{secondhand} codes; because these activities involve creating apparatus, we labeled them \textbf{fabrication-oriented}.}
\item \drdf{The five remaining activities were not characterized by obvious coding patterns, but they almost all relate to propagation of scientific knowledge to and from the project team; we labeled them \textbf{propagation-oriented}.}
\end{itemize}
Most activities were identified as \drdf{\emph{definitely}} or \emph{might be a feature of experimental physics research} by all participants. Six activities---including \drdf{three} fabrication-oriented activities---were identified as \emph{not a feature} by one or two participants. For all but one activity, a majority of participants referred to \emph{firsthand experiences}. \drdf{The exception} was writing code to simulate results, a fabrication-oriented activity for which only about a third of participants referred to \emph{firsthand experiences}.

\textbf{Execution-oriented activities} were more frequently described as \emph{necessary} than those in other categories, and, on the Likert-style survey, they were identified as \emph{definitely a feature of experimental physics research} by almost all participants.
Compared to other activities, troubleshooting was most frequently described as \emph{necessary}. Consider the following \drdf{response from a student we call Brittany}:
\begin{itemize}
\item[] ``Troubleshooting problems with equipment. Huge. We had so many problems. I said `definitely a feature' because troubleshooting problems is one of the biggest parts of lab projects that I've done, no matter what project it was. You know, you have an idea, and of course it's not going to come off exactly right. So that's big. And then if you have something running for a long time, it won't stay running at the top of its game for its entire life. You're going to have to fix it sometimes.'' 
\end{itemize}
Brittany described troubleshooting as ``huge'' and ``one of the biggest parts of lab projects,'' and she referred to her \emph{firsthand experience} encountering problems on her final project and all other projects. Brittany framed the need to troubleshoot as an inevitable part of experimentation (cf. Ref.~\cite{Dounas-Frazer2016}), indicating that she viewed troubleshooting problems with equipment to to be \emph{necessary} for experimental physics.

Compared to other activities, troubleshooting and maintaining a lab notebook were most frequently described as \emph{definitely a feature of experimental physics research}. Consider the following \drdf{response from a student we call Ashley}:
\begin{itemize}
\item[] ``[Maintaining a lab notebook] is typically a very big part of research because you need to show proof. Especially, like, if you're in industry, you definitely need proof that you did things at certain times so that competing companies know that you did do it first. Or, if you're in academia, you need that notebook for when you graduate from your group, for people coming in to use your work, and having a notebook that's clear is always helpful. Even in writing this [lab report], there's things, as I run through the notebook, that I hadn't realized group members had done that I needed to incorporate, so that's important.'' 
\end{itemize}
Ashley described lab notebooks as ``very big,'' ``always helpful,'' and ``important'' in corporate, academic, and educational settings, and she referred to her \emph{firsthand experience} relying on a notebook when writing a report for her project. Ashely said that notebooks are needed to provide timestamped evidence of milestones and to facilitate knowledge transfer between group members, indicating that she viewed maintaining a lab notebook to be \emph{necessary} for experimental physics.

\textbf{Interpersonal activities} were identified as \emph{definitely a feature of experimental physics research} by a majority of participants, and they accounted for almost all instances in which participants discussed the \emph{status-dependent} nature of an activity.
Compared to other activities, asking a supervisor for help was most frequently described as \emph{status-dependent}. Consider the following \drdf{response from a student we call Michael}:
\begin{itemize}
\item[] ``Asking a supervisor for help I said was just maybe. Some people could be way better at this than I am. And so therefore they know exactly what they're supposed to do, and after the supervisor tells them right at the beginning, `You're going to want to do this work,' they can just go. And there have been times at the lab where [the project team] have gone a week without talking to our professor about [the project] because we have a very clear idea of where we want to go. So---and also, if you were at the top of your field, doing research that no one has done anything like before, then you may not have a supervisor to ask for help.'' 
\end{itemize}
Michael referred to his \emph{firsthand experience} working on his project without input from his professor to illustrate his view that it is possible to conduct research without asking a supervisor for help.  According to Michael, people with high research competence are unlikely to ask a supervisor for help, and someone who is ``at the top'' of their field may not be able to do so, indicating that he viewed asking a supervisor for help as a \emph{status-dependent} aspect of experimentation.

\textbf{Fabrication-oriented activities} were identified as \emph{might be a feature of experimental physics research} by a majority of participants, and they accounted for almost all instances in which participants discussed the \emph{resource-dependent} nature of an activity or referred to their \emph{secondhand experiences}.
Compared to all other activities, fabrication was most frequently described as \emph{resource-dependent}. Consider the following \drdf{response from a student we call Brandon}:
\begin{itemize}
\item[] ``Fabricating parts I said `might be' because you might be able to just find a company that's manufacturing the thing, the equipment you need, so you might be able to just buy it. Or, if you can't find it, or you find something that's close but not exactly---you might have to kind of edit it or go into a shop and actually build it. Like, I know one lab group has done that a whole lot.'' 
\end{itemize}
Brandon referred to a \emph{secondhand experience} in which a group built parts of their apparatus to illustrate his view that commercial availability of equipment can inform whether or not experimentation involves fabrication. Thus, Brandon's view is consistent with the notion that fabricating parts or materials is a \emph{resource-dependent} aspect of experimentation.

\textbf{Propagation-oriented activities} were identified as \emph{definitely a feature of experimental physics} by a majority of participants. However, we did not notice other obvious patterns in code assignments. For example, while reading scientific papers was described as \emph{necessary} by a majority of participants, presenting results orally was more frequently described as \emph{not necessary} than any other activity. Consider the following \drdf{response from a student we call Logan}:
\begin{itemize}
\item[] ``[Presenting orally] can be helpful for explaining your results to other people, but I wouldn't say it's absolutely essential since you can always just publish \ldots'' 
\end{itemize}
Logan reasoned that, because oral presentations are not the only mechanism for propagating research findings, presenting results orally is \emph{not necessary} for experimental physics.

\vspace{-10pt}
\section{Discussion and limitations}
{In response to research question Q1,} almost all participants viewed the activities in Table~\ref{table:performances} to be constitutive performances of experimental physics practice, sometimes conditionally. Execution-oriented activities were often viewed as necessary for experimentation, whereas interpersonal and fabrication-oriented activities were viewed \drdf{as conditional} aspects of experimental physics, depending on the novice or expert status of team members and the availability of commercial apparatus. Participants' views on propagation-oriented activities \drdf{were mixed}, but not because propagation of knowledge itself was viewed as disconnected from experimental physics practice. Rather, participants acknowledged a variety of avenues through which research findings can be shared \drdf{with others}, some of which may be prioritized over others.

{In response to Q2}, participants regularly referred to their firsthand experiences working on projects to illustrate or justify their views about the role of an activity in experimentation. Because some projects did not involve building parts or writing code, participants' views about fabrication-oriented activities were often informed by secondhand experiences {and perceived to be} conditional aspects of experimentation. \drdf{Thus, participants' views about experimental physics were shaped by their own experiences working on multiweek projects and their perceptions of their peers' experiences.}

\drdf{Our study design constrains} the generalizability of our findings. \drdf{We} investigated students' views about only a subset of activities that are relevant to experimentation and that arose in a particular educational context. Moreover, white men are more overrepresented in the courses we studied than among physics bachelor's degree recipients in the United States, which likely {contributes to} homogeneity of the views reported here. Indeed, as measured by E-CLASS, on average, women's and men's views about experimental physics differ in some respects, and nonbinary people's views have not been explored~\cite{Wilcox2016a}. \drdf{Hence, we cannot achieve} probabilistic generalization to all students or all lab courses. Instead, we strive for theoretical generalization~\cite{Eisenhart2009}, which, for us, involves {inferring some plausible} mechanisms through which {projects may influence} students' views about experimental physics.

One such mechanism is repeated firsthand experience with an activity, similar to Brittany's experiences with troubleshooting. Brittany's view of troubleshooting as inevitable is consistent with a common stance among electronics lab instructors: troubleshooting does not need to be explicitly taught or assessed in lab courses because `nothing works the first time,' and the need to troubleshoot arises organically~\cite{Dounas-Frazer2017a,Dounas-Frazer2016}. {Irving and Sayre~\cite{Irving2014} argued} that students are accountable to a different kind of physics knowledge in lab courses than in other courses. It is plausible that students come to view troubleshooting as inherent to experimentation because technical problems are frequent, and students are implicitly accountable for learning how to troubleshoot them. 

\drdf{Engaging groups in unique projects may be another mechanism that shapes students' views about experimentation.} {Irving and Sayre~\cite{Irving2014} argued that students develop experiment-specific expertise \drdf{when groups} work on distinct projects at the same time. Similarly, \drdf{we find that} students notice experiment-specific approaches to experimentation. Comparing and contrasting their own approaches to those of others could give rise to combinations of firsthand and secondhand experiences through which students develop nuanced ideas about experimental physics as comprising context-dependent combinations of performances. Future work could explore the impact of working on different projects at the same time on students' views about experimental physics.

Although we did not engage with Ford's~\cite{Ford2015} idea that practice comprises performances that are connected and whose purpose is to explain nature better, we see hints that some participants viewed some activities as purposeful. Ashley recognized that maintaining a lab notebook plays multiple roles in the experimental process, in alignment with what Hoehn and Lewandowski refer to as `writing as professionalization'~\cite{Hoehn2020}. Ashley's perception of lab notebooks as ``always helpful'' is in contrast to work by Stanley and Lewandowski~\cite{Stanley2016b}, who found that many graduate students do not view lab notebooks as a beneficial part of their undergraduate lab courses. However, evaluating student responses like Ashley's using Ford's notion of purposefulness is beyond the scope of the present work. Future work could explore how final projects in lab course support students to view performances as connected or purposeful \drdf{(cf. Ref.~\cite{Quan2018})}.

\vspace{-7pt}
\acknowledgments{\redact{ S.\ Lavender and L.\ Torres helped with undergraduate research logistics, C.\ Hoyt helped collect data, L.\ Kiepura helped transcribe interviews, and the WWU PER Group (especially A.\ Boudreaux, R.\ Barber DeGraaff, and T.\ L\^e) helped interpret results. This material is based on work supported by the NSF under Grant Nos.\ 1726045, 1323101, 1734006, and 1208930, and by the Washington NASA Space Grant Consortium under Grant No.\ NNX14AR60A.}}

\clearpage

\bibliography{practice_database_2019} 

\end{document}